\documentclass[%
 aip,
 jap,
 amsmath,amssymb,
preprint
]{revtex4-1}

\usepackage{graphicx}
\usepackage{epstopdf}
\usepackage{hyperref}

\begin{document}

\preprint{JAP}

\title{Ground-state magnetic phase diagram of bow-tie graphene nanoflakes \\in external magnetic field}

\author{Karol~Sza{\l}owski}
\email{kszalowski@uni.lodz.pl; kszalowski@wp.pl}
\affiliation{Department of Solid State Physics, Faculty of Physics and Applied Informatics, University of {\L}\'{o}d\'{z}, ul. Pomorska 149/153, 90-236 {\L}\'{o}d\'{z}, Poland}

\date{\today}

\begin{abstract}
The magnetic phase diagram of a ground state is studied theoretically for graphene nanoflakes of bow-tie shape and various size in external in-plane magnetic field. The tight-binding Hamiltonian supplemented with Hubbard term is used to model the electronic structure of the systems in question. The existence of the antiferromagnetic phase with magnetic moments localized at the sides of the bow-tie is found for low field and a field-induced spin-flip transition to ferromagnetic state is predicted to occur in charge-undoped structures. For small nanoflake doped with a single charge carrier the low-field phase is ferrimagnetic and a metamagnetic transition to ferromagnetic ordering can be forced by the field. The critical field is found to decrease with increasing size of the nanoflake. The influence of diagonal and off-diagonal disorder on the mentioned magnetic properties is studied. The effect of off-diagonal disorder is found to be more important than this of diagonal disorder, leading to significantly widened distribution of critical fields for disordered population of nanoflakes.

\end{abstract}




\maketitle

\section{Introduction}

Spintronics is hoped to replace and significantly extend the possibilities of information processing based on charge degrees of freedom for electrons. An applicational potential of graphene\cite{Novoselov1,Novoselov2} - an unique two-dimensional novel material - is boosted by developing spintronic devices based on its magnetic properties\cite{Spin2013,Seneor2013,Rycerz2007}. In the context of intriguing magnetic characteristics, various graphene-based nanostructures are invoked, mainly to mention graphene nanoflakes (GNFs) or quantum dots, for which the ability of shaping the edge and designing the electronic structure allows to reach desirable properties \cite{Snook,Snook2,Ritter2009,Ezawa2007,EzawaNJP,Ezawa2009,Ezawa2010,Ominato2013,Ominato2013b,Koshino,Zhou2012,Szafran,Rossier2007,Wang2008,Wang2008b,Wang2009,Potasz2010,Potasz2012,Guclu2009,Guclu2013,Szalowski1,Szalowski2,SzalowskiAPPA}. Numerous applications of these zero-dimensional graphene-based structures within an emerging field of spin electronics are suggested in the recent theoretical works \cite{Weymann2,Weymann,Ezawa2010,Ezawa2007,Ezawa2009,Ezawa2010,EzawaNJP,Wang2009,Kang2012,Silva2010,Wang2008,Li2012,Guclu2011,Sheng2010,Candini}. Among various graphene nanoflakes, particularly those of triangular shape and zigzag edge seem promising and attract the attention \cite{Wang2008,Silva2010,Ezawa2012,Potasz2010,Potasz2012,Guclu2009,Guclu2013,Zhou2011,Guo2013,Jaworowski2013,Szalowski2,Yoneda2009,SzalowskiAPPA}. This is due to presence of a shell of zero-energy states owing to imbalance in number of atoms belonging to two interpenetrating sublattices\cite{Fajtlowicz,Yazyev2010}. Such states become localized at the zigzag edge (which effect is confirmed for various graphene structures \cite{Wakabayashi1996,Klusek2000,Klusek2001,Klusek2005b,Kobayashi2005}) and result in spin polarization of the flake edge \cite{Lieb,Wang2008,Rossier2007,Potasz2010,Guclu2009}. However, also bow-tie GNFs, which can be considered to some extent as structures composed of two triangular flakes, constitute an interesting class of graphene quantum dots\cite{Wang2009}. Bow-tie GNFs were predicted to show magnetic moments localized mainly at their sides and oriented antiparallel in both halves of the nanostructure \cite{Wang2009,Yazyev2010,Potasz2010,Ijas2010,Yoneda2009,Kang2012,Sheng2013,Zhou2013}. Their potential spintronic applications were studied in Refs.~\onlinecite{Wang2009,Kang2012}. 

One of the goals in theoretical description of graphene magnetic nanostrucutures is characterization of the influence of external electric and magnetic fields on their properties. The recent studies concern both zero-dimensional graphene structures \cite{Guclu2011,Guclu2013,Potasz2012,Zhou2013,Lu2012,Agapito2010,Rycerz2010,PotaszAPPA} and systems of higher dimensionality (e.g. \cite{Jaskolski}). In particular, the external electric field of gates has been very recently predicted to switch between antiferromagnetic and nonmagnetic state in bow-tie graphene nanoflakes\cite{Zhou2013}. This encourages interest in influence of the external field on the phase diagram of graphene nanostructures.

The aim of our work is to investigate the effect of the external magnetic field on the ground state magnetic phase diagram of a bowtie-shaped GNF. In order to test the robustness of the predicted behaviour, we also study the influence of the disorder on the predicted properties.

\section{Theoretical model}

\begin{figure}[t]
  \begin{center}
   \includegraphics[scale=0.3]{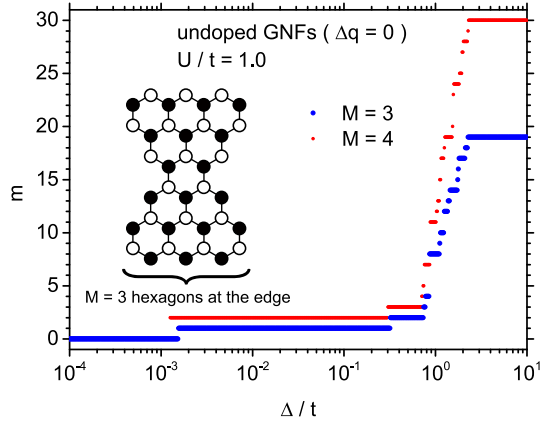}
  \end{center}
   \caption{\label{fig:fig0}Dependence of total magnetizations of two undoped GNFs ($M=3$ and $M=4$) on the normalized external field. In the inset a schematic view of a bow-tie GNF with $M=3$ hexagons at the side is presented, with empty/filled circles representing carbon atoms belonging to two sublattices.}
\end{figure}

The subject of interest is graphene nanoflakes of bow-tie shape (see inset in Fig.~\ref{fig:fig0}), with $M$ hexagons forming each side of a bow-tie. The nanoflake is assumed to contain $N=2M^2+8M-5$ carbon atoms belonging to two interpenetrating sublattices. Moreover, it contains $N+\Delta q$ electrons on $p^{z}$ orbitals, which are crucial for the electronic structure of graphene. The case of $\Delta q=0$ corresponds to charge-neutral, undoped structure, while $\Delta q=\pm 1$ denotes doping with a single chare carrier (electron/hole).
In order to describe the electronic structure of the GNFs we use the following Hamiltonian:
\begin{eqnarray}
\label{eq:eq1}
\mathcal{H}_{0}&=&\sum_{i,\sigma}^{}{\epsilon_{i}\,c^{\dagger}_{i,\sigma}c^{}_{i,\sigma}}-\sum_{\left\langle i,j\right\rangle ,\sigma}^{}{t_{ij}\,\left(c^{\dagger}_{i,\sigma}c^{}_{j,\sigma}+c^{\dagger}_{j,\sigma}c^{}_{i,\sigma}\right)}\nonumber\\&+&U\sum_{i}^{}{\left(n_{i,\uparrow}\left\langle n_{i,\downarrow} \right\rangle+n_{i,\downarrow}\left\langle n_{i,\uparrow} \right\rangle\right)}\nonumber\\&-&U\sum_{i}^{}{\left\langle n_{i,\uparrow} \right\rangle\left\langle n_{i,\downarrow} \right\rangle}+\frac{\Delta}{2} \sum_{i}^{}{\left(c^{\dagger}_{i,\uparrow}c^{}_{i,\uparrow}-c^{\dagger}_{i,\downarrow}c^{}_{i,\downarrow}\right)}.
\end{eqnarray}
It consists of a tight-binding part with nearest-neighbour hopping energy equal to $t_{ij}=t$ (in absence of any disorder) supplemented with Hubbard on-site term, which is subject to Mean Field Approximation (MFA). The on-site energy $\epsilon_{i}$ is set to 0 unless a diagonal disorder is included. Let us state that such a model is quite commonly used for description of magnetic properties of graphene and its derivative nanostructures (e.g. \cite{Rossier2007,Yazyev2010}). It is worth noticing that the value of the Hubbard $U$ parameter here is not a true on-site repulsion energy, but instead it is an effective parameter taking into consideration coulombic interactions between electrons at different sites (see Ref. \cite{Schuler2013}). Therefore, its value is significantly reduced with respect to on-site value and we accept $U/t=1.0$ in our calculations. Moreover, the long-range coulombic behaviour has been recently found to be suppressed in GNFs of the shape we consider here \cite{Sheng2013}. Moreover, the MFA treatment of Hubbard model is known to predict the total energy values which are consistent with the results of either exact diagonalization or Quantum Monte Carlo simulations for $U/t<2.0$ \cite{Feldner2010}, which is the case in graphene nanostructures. The importance of Hubbard term has been also observed for indirect Ruderman-Kittel-Kasuya-Yosida coupling in graphene and its nanostructures \cite{Annika2010,Szalowski1,Szalowski1,SzalowskiAPPA}.

The external in-plane magnetic field $H$ is included in the Hamiltonian by means of a Zeeman term, in which $\Delta=g\mu_{\rm B}H$ is the Zeeman splitting energy, which parameter is used in further considerations to parametrize the external field. We emphasize that our interest is limited to in-plane field. Therefore, the Peierls substitution\cite{Hofstadter}, consisting in modification of the hopping integral by complex phase factor dependent on field vector potential, is not used here, since only the field perpendicular to the plane modifies the hopping integrals.

The total ground-state energy of the charge carriers is determined in a self-consistent procedure of diagonalization of the Hamiltonian \ref{eq:eq1} for a fixed number of charge carriers $N+\Delta q$ (for which a procedure LAPACK package is utilized\cite{lapack}). Summation of $N+\Delta q$ least eigenvalues of the Hamiltonian allows to calculate the total energy of the ground state. In addition, the knowledge of the self-consistent distribution of spin-up and spin-down electrons over the lattice sites allows to characterize the magnetic moment of the nanoflake and its distribution.

\begin{figure}[t]
  \begin{center}
   \includegraphics[scale=0.3]{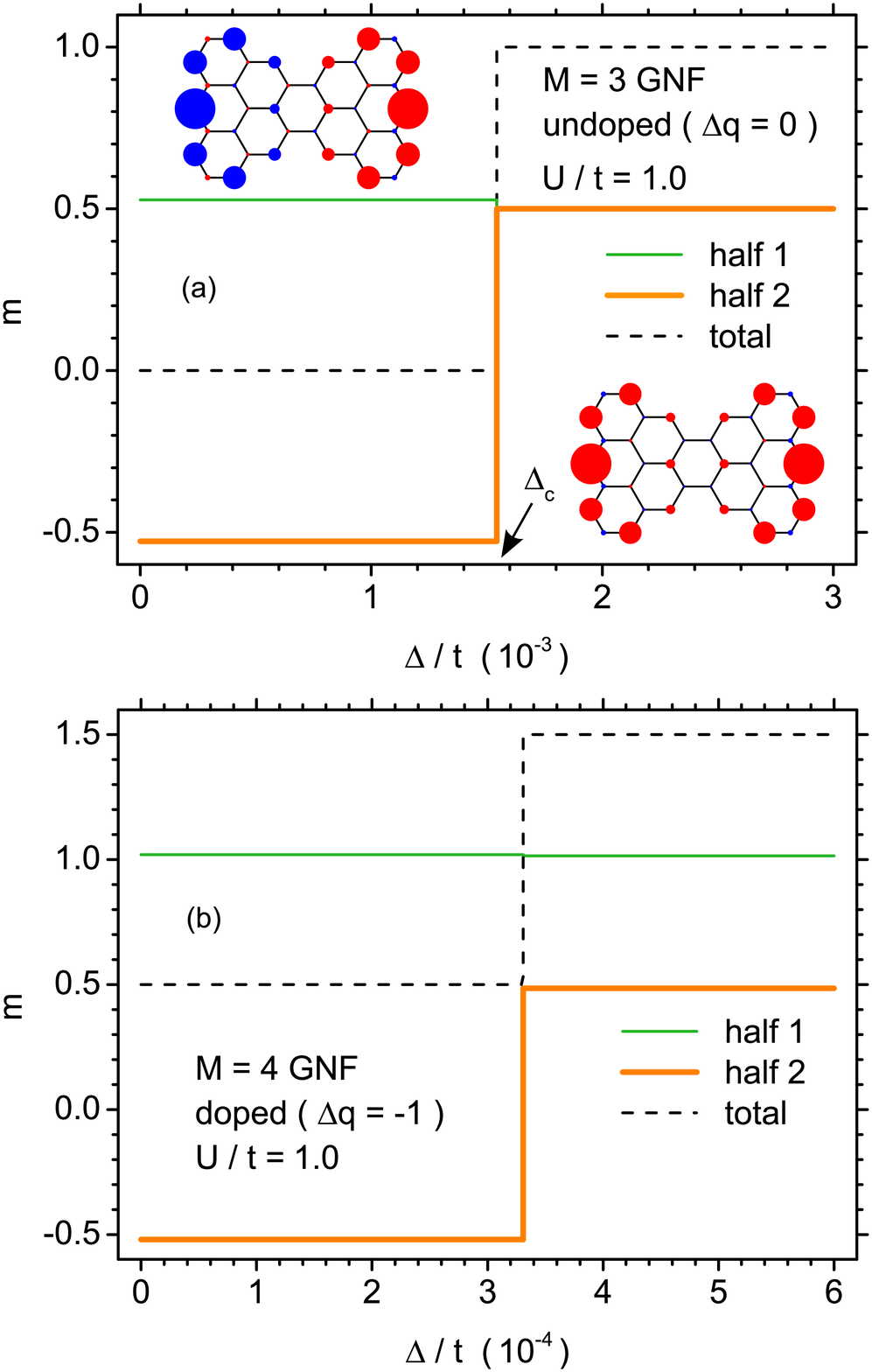}
  \end{center}
   \caption{\label{fig:fig4}Dependence of magnetizations of GNF halves and total magnetization of a GNF on the normalized external field, for charge-undoped GNF with $M=3$ (a) and $M=4$ GNF doped with a single hole (b). In the inset the distribution of magnetic moment on the lattice sites of GNF is shown for weak and for strong external field.}
\end{figure}

\begin{figure}[t]
  \begin{center}
   \includegraphics[scale=0.3]{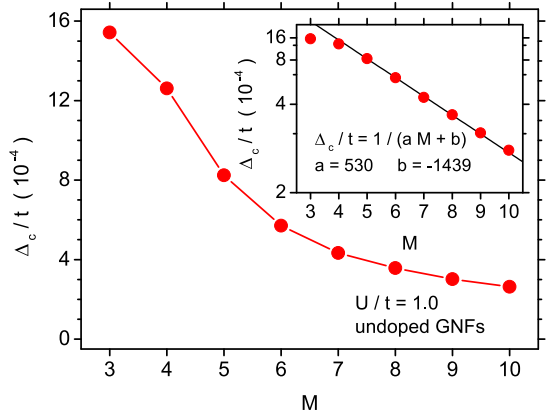}
  \end{center}
   \caption{\label{fig:fig1}Dependence of the normalized critical field for spin-flip transition on GNF size (number of hexagons at each side of a bow-tie). In the inset the same dependence is shown in inverse scale, with the empirical function $\Delta_{c}=1/\left(aM+b\right)$ fitted. }
\end{figure}

\begin{figure}[t]
  \begin{center}
   \includegraphics[scale=0.3]{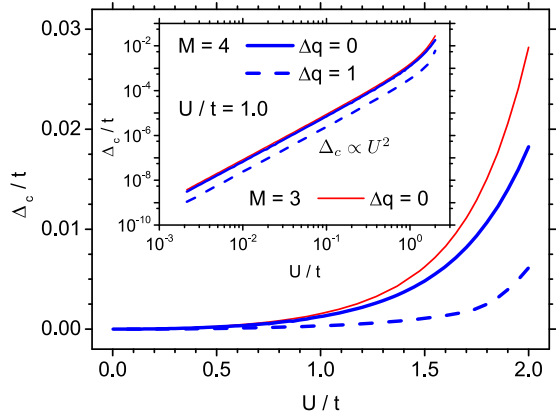}
  \end{center}
   \caption{\label{fig:fig3}Normalized critical field as a function of the Hubbard on-site parameter $U$. For undoped structures the critical field corresponds to spin-flip transition between antiferromagnetic and ferromagnetic state while for a structure doped with a single charge carrier the transition is between ferrimagnetic and ferromagnetic state. The inset presents the data in doubly logarithmic scale.}
\end{figure}

\begin{figure}[t]
  \begin{center}
   \includegraphics[scale=0.3]{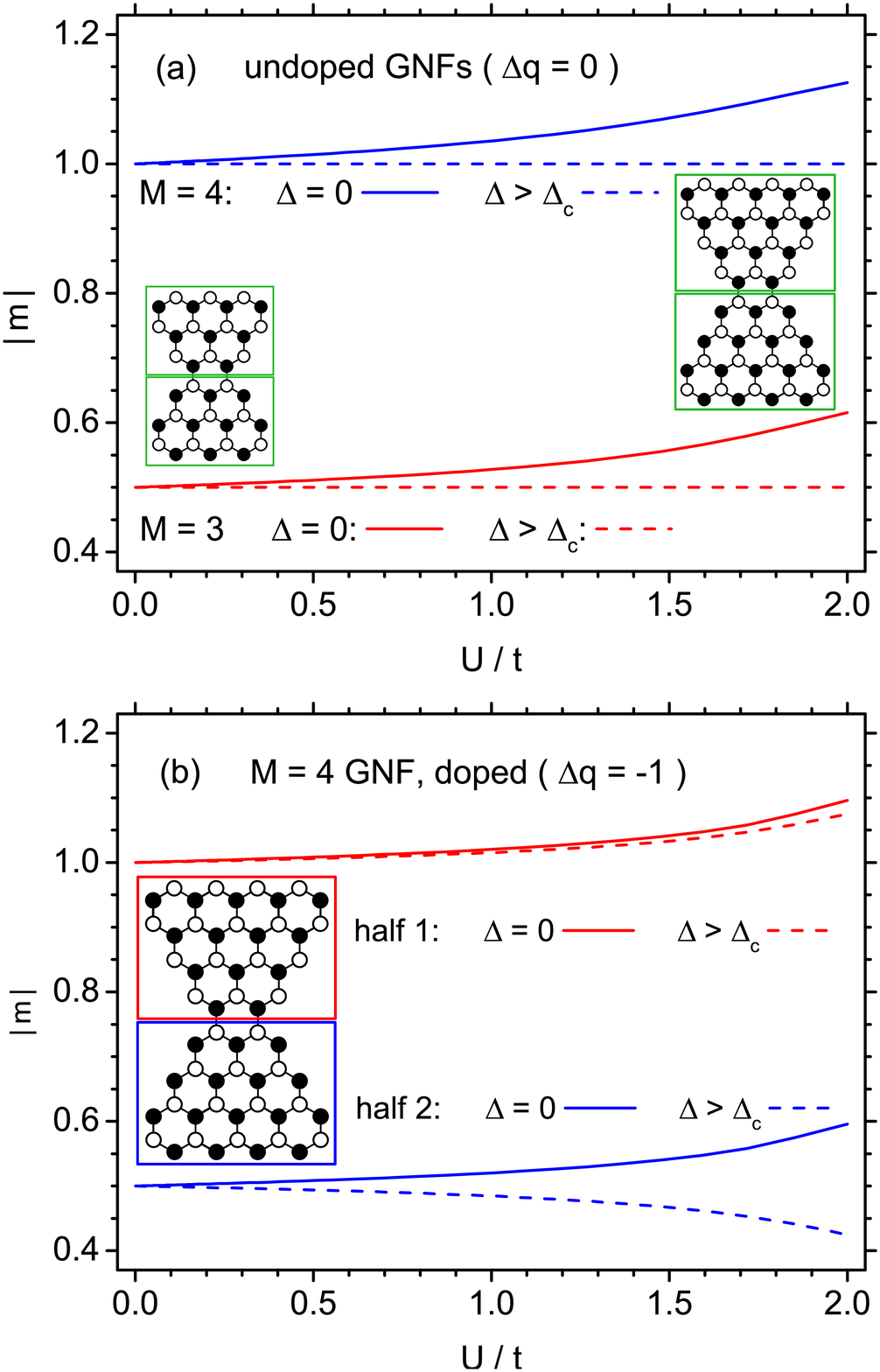}
  \end{center}
   \caption{\label{fig:fig2}(a) Dependence of an absolute value of magnetization of each half of the GNF for undoped $M=3$ and $M=4$ structures on the Hubbard $U$ parameter. (b) Dependence of absolute values of magnetization of both halves of the $M=4$ GNF doped with a single hole on the Hubbard $U$ parameter. Solid (dashed) lines correspond to field lower (higher) than the critical field.}
\end{figure}

\begin{figure}[t]
  \begin{center}
   \includegraphics[scale=0.3]{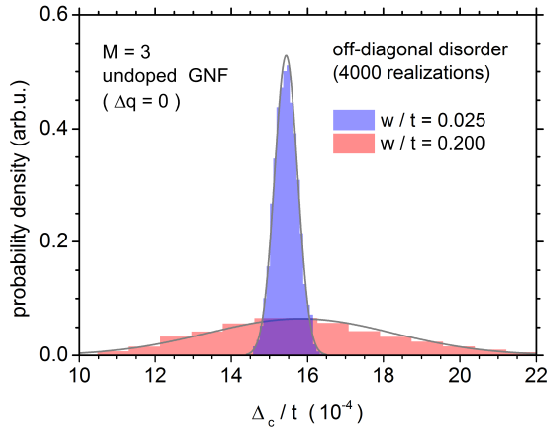}
  \end{center}
   \caption{\label{fig:fig5}Distribution of normalized critical field values for an undoped $M=3$ GNF for 4000 realizations of off-diagonal disorder. Two strengths of disorder were used, characterized by half-widths of uniform distribution for hopping integral equal to $w/t=0.025$ and $0.20$. Solid lines denote fitted normal distributions.}
\end{figure}

\begin{figure}[t]
  \begin{center}
   \includegraphics[scale=0.3]{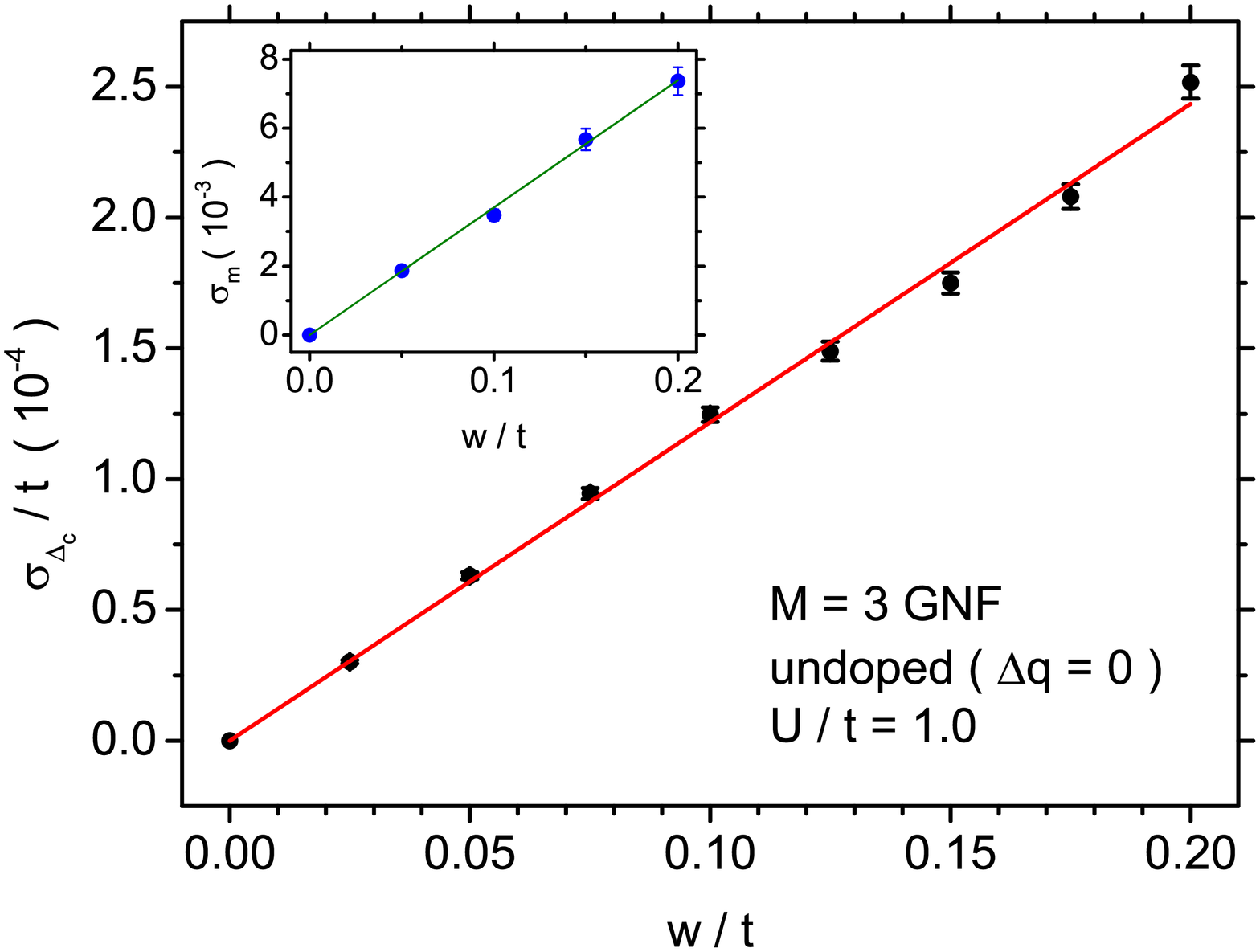}
  \end{center}
   \caption{\label{fig:fig6}Dependence of the standard deviation for critical field distribution for $M=3$ undoped GNF on the off-diagonal disorder strength (half-width of uniform distribution for hopping integrals). The inset presents analogous dependence of the standard deviation of low-field magnetization magnitude of each half of a GNF. Solid lines show linear dependencies fitted to the data.}
\end{figure}

\begin{figure}[t]
  \begin{center}
   \includegraphics[scale=0.3]{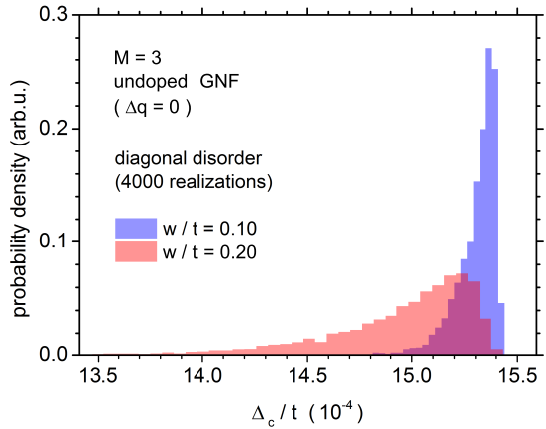}
  \end{center}
   \caption{\label{fig:fig7}Distribution of normalized critical field values for an undoped $M=3$ GNF for 4000 realizations of diagonal disorder. Two strengths of disorder were used, characterized by half-widths of uniform distribution for hopping integral equal to $w/t=0.10$ and $0.20$.}
\end{figure}

\begin{figure}[t]
  \begin{center}
   \includegraphics[scale=0.3]{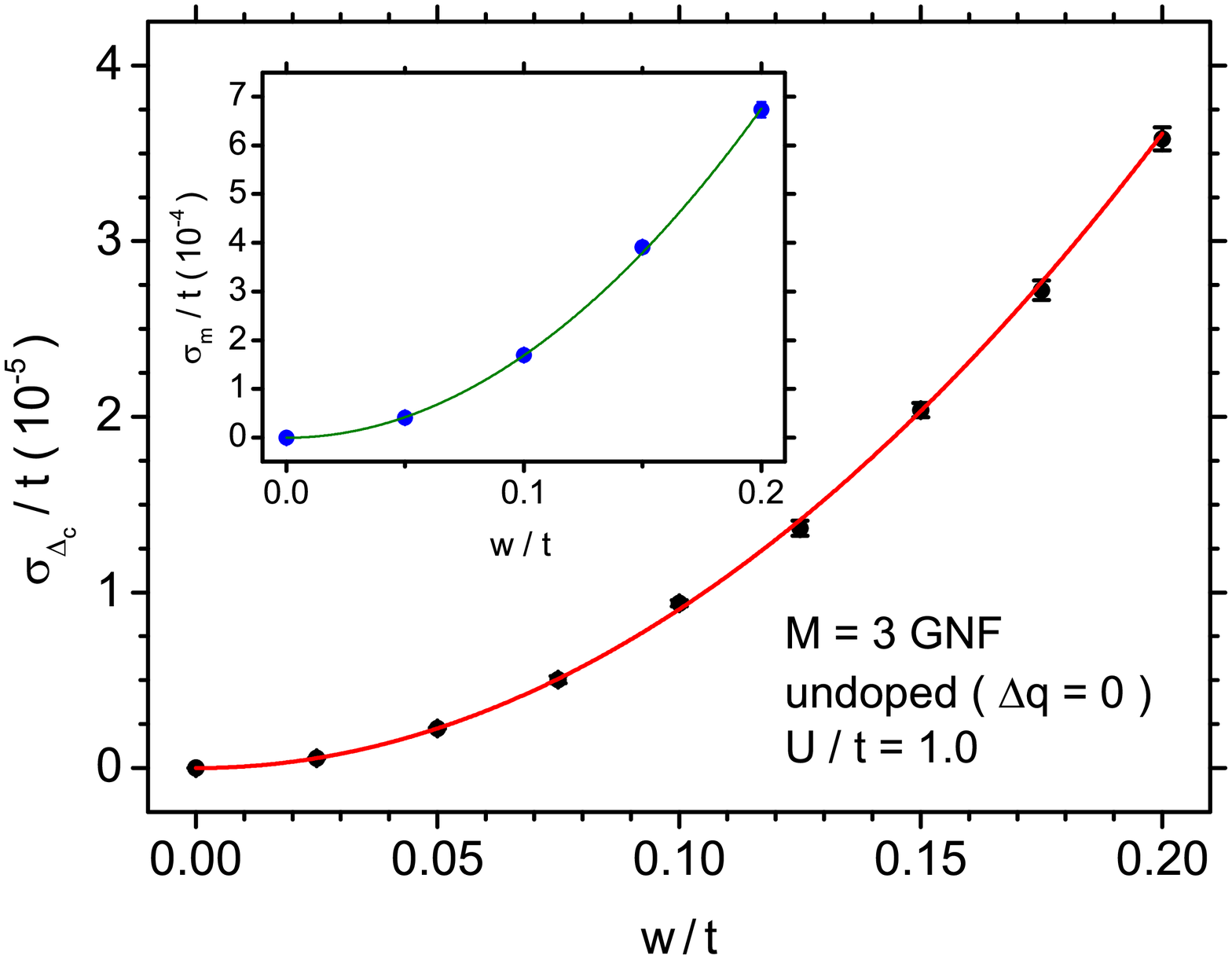}
  \end{center}
   \caption{\label{fig:fig8}Dependence of the standard deviation for critical field distribution for $M=3$ undoped GNF on the diagonal disorder strength (half-width of uniform distribution for on-site energies). The inset presents analogous dependence of the standard deviation of low-field magnetization magnitude of each half of a GNF. Solid lines show quadratic dependencies fitted to the data.}
\end{figure}

\section{Results}

In order to construct a ground-state phase diagram of the GNF in external magnetic field, we study first the dependence of the GNF magnetization on the field. The total magnetization of two small undoped GNFs, with $M=3$ and $M=4$, is plotted as a function of the normalized external field in Fig.~\ref{fig:fig0}. In principle it is visible that the dependence is composed of a series of magnetization plateaus with discontinuous field-induced changes between them. The field range corresponding to particular plateaus varies strongly in width. In particular, for quite wide low field range (including zero field) the total magnetization is equal to zero. The next plateau (ferromagnetic state with low spin) is also considerably robust against the field increase. On the contrary, the next plateaus associated with higher magnetic moments correspond to much narrower ranges of external field and a series of discontinuous transitions occurs in a limited range of $\Delta$ causing the fast increase of the total spin up to the saturation value of $N/2$, where $N$ is the number of electrons in undoped GNF (equal to the number of the carbon atoms). We notice also that the external field corresponding to the transition between second and third plateau occurs at a field about 2 orders of magnitude higher that the transition between first and second plateau. Therefore, the most interesting range is certainly the zero-spin state and the next ferromagnetic state with low magnetic moment. In the following part of our considerations we will focus our attention on this range.  

The detailed results for undoped $M=3$ GNF concerning the magnetization changes with the field for the mostly interesting range are presented in Fig.~\ref{fig:fig4}(a). The total GNF magnetization value is plotted together with the magnetic moments of both halves of a GNF. It is visible that for low (or zero) external field, the GNF is polarized antiferromagnetically and the magnetization values do not vary with the field. At certain strength of the critical external field $\Delta_{c}$, a spin-flip transition occurs between antiparallel and parallel orientation of magnetic moments of both halves of a GNF. Then the further increase of the field does not change the magnetizations. It is also visible that after the spin-flip transition the absolute values of the magnetizations are slightly reduced. The distribution of magnetic moments over the carbon lattice sites for a $M=3$ GNF below and above $\Delta_{c}$ are depicted in the insets in Fig.~\ref{fig:fig4}(a), showing that the magnetic moment is dominantly concentrated near the edges with largest values close to the sides of a bow-tie. In the insets the different colours indicate opposite orientations of magnetic moments. An analogous plot is presented in Fig.~\ref{fig:fig4}(b) for the case of a $M=4$ GNF doped with a single hole ($\Delta q=-1$). In such a situation we deal with an occurrence of a kind of a metamagnetic transition, with low-field ferrimagnetic state and high-field ferromagnetic state (since the magnetization magnitudes of both GNF halves are unequal for doped nanonstructure). 

It is of particular interest to study the influence of GNF size on the value of the critical field $\Delta_{c}$ for spin-flip transition between antiferromagnetic and ferromagnetic state of undoped GNFs. Such a dependence is plotted in Fig.~\ref{fig:fig1}, in which $\Delta_{c}$ is presented as a function of the number $M$ of hexagons forming each edge of a bow-tie (see schematic inset). A non-linear decrease of the critical field with increasing GNF size is observable. In the inset plot $\Delta_{c}$ is presented as dependent on $M$ in inverse scale. Such a plot allows to notice that for sufficiently large GNFs, namely for $M>4$, the results of calculations can be well fitted with the dependence $\Delta_{c}/t=1/\left(aM+b\right)$ which is plotted with a solid straight line in the inset and the $a$ and $b$ parameter values are given there. 

The importance of the Hubbard parameter $U$ can be analysed on the basis of the Fig.~\ref{fig:fig3}, which presents the critical field as a function of $U$. Let us remind that for undoped cases ($\Delta q=0$) $\Delta_{c}$ corresponds to transition between antiferromagnetic and ferromagnetic state, while for $\Delta q=-1$ it separates ferrimagnetic and ferromagnetic ordering. We should emphasize that for $U=0$ the ferro- and antiferromagnetic (or ferrimagnetic) state are degenerate (have the same energy) and thus the critical field $\Delta_{c}\to 0$ when $U\to 0$. In the inset in Fig.~\ref{fig:fig3} the same data are plotted in doubly logarithmic scale, which allows to notice that the critical field is proportional to $U^2$ provided that $U$ is not too large (below $U/t\simeq 1$). Therefore, the physically relevant regime for GNFs is located slightly above this threshold and $\Delta_{c}$ rises faster than quadratically in this range.

It is also interesting to study the dependence of low- and high-field magnetizations on the selection of Hubbard parameter $U$. For the case of undoped nanoflakes with $M=3$ and $M=4$, such a dependence is illustrated in Fig.~\ref{fig:fig2}(a). The plotted values are the absolute values of the magnetization of each half of a GNF, either for low field $\Delta\to 0$ or high field exceeding $\Delta_{c}$. It is evident that below the critical field $\Delta_{c}$, the magnetizations rise slightly with $U$, while for $\Delta>\Delta_{c}$ they remain insensitive to the Hubbard parameter value (which follows from the fact that the magnetization is saturated within the plateau ranging up to considerably high fields). Let us state that the magnetization directions for $\Delta<\Delta_{c}$ are antiparallel, so that the total magnetization remains 0. The situation is somehow different for a doped $M=4$ GNF (as illustrated in Fig.~\ref{fig:fig2}(b)). There, magnetizations magnitudes of both halves of a GNF are unequal. As long as $\Delta<\Delta_{c}$, both of them rise with increasing field. However, the situation changes above the critical field, when the larger of magnetizations tends to increase with increasing field, while the other one exhibits quite the opposite tendency. In both regimes the total magnetization of a nanoflake is constant.

In order to make a step towards estimating the robustness of the described effects against disorder, we performed additional calculations in which we included the presence of either diagonal or off-diagonal disorder in the Hamiltonian (Eq.~\ref{eq:eq1}). 

In the case of an off-diagonal disorder, the hopping integrals in Hamiltonian (Eq.(\ref{eq:eq1})) were expressed as $t_{ij}=t+\Delta t_{ij}$, where $\Delta t_{ij}\in \left[-w,w\right]$ are random variables taken from a uniform distribution of half-width $w$ centered at 0. Various values of distribution width $w$ ranging up to $w/t=0.20$ were considered. In the calculations, the values of $\Delta t_{ij}$ were generated using a random number generator described in Ref.~\cite{MersenneTwister}. Let us observe that such a model of disorder corresponds to a disorder in bond lengths between carbon atoms, since the hopping integral depends on the bond length $a_{ij}$ like $t_{ij}=t_0 e^{-\beta\left(\frac{a_{ij}}{a}-1\right)}$ \cite{Ribeiro2009}. Therefore, $\Delta t_{ij}/t\simeq -\beta \Delta a_{ij}/a$ with $\beta\simeq 3$\cite{Ribeiro2009,Pereira2009} is valid for small bond deformations. The largest used value of $w/t=0.20$ corresponds roughly to maximum relative bond deformation of $~7\%$. 

In Fig.~\ref{fig:fig5} we plot a histogram of critical field values obtained for undoped $M=3$ GNFs, for two values of disorder strength: $w/t=0.025$ and $0.20$. For each case the population of 4000 GNFs with random hopping integrals was examined. It is visible that both distributions follow the normal probability distribution (fitted solid lines in Fig.~\ref{fig:fig5}). Its dispersion is quite low for $w/t=0.025$ and becomes very significant for $w/t=0.20$, when the distribution becomes rather wide. On the other hand, the average value of critical fields is only very weakly sensitive to disorder, as the maxima of both distributions show a slight shift. In order to illustrate the influence of the disorder on the critical field distribution width, we plotted the standard deviation of critical field distribution $\sigma_{\Delta_c}$ as a function of $w$ in Fig.~\ref{fig:fig6}. It is visible that the distribution dispersion is a linear function of hopping integral distribution width $w$ in the whole studied range up to $w/t=0.20$ and for the highest value of $w/t=0.20$ the $\sigma_{\Delta_{c}}$ is a significant fraction of $\Delta_c$, approximately 16$\%$. Therefore, the bond disorder has a significant influence on the critical field in GNFs. In the inset in Fig.~\ref{fig:fig6} we also depicted the standard deviation of the probability distribution for the magnetization of each half of a$M=3$ GNF for $\Delta<\Delta_{c}$ as a function of $w$. While $\sigma_{m}$ is also proportional to $w$, yet the magnetization distribution width is negligible in comparison with the average value of magnetization (which is close to 0.5 for that case, so that the relative half-width is less than 2$\%$).

For the case of a spin-independent diagonal disorder, we treat the on-site energy $\epsilon_{i}$ in the Hamiltonian (Eq.(\ref{eq:eq1})) as a random variable taken form the uniform distribution $\epsilon_{i}\in\left[-w,w\right]$ and obtained in the same way as for off-diagonal disorder. As it is visible in Fig.~\ref{fig:fig7}, where a histogram of critical field values is presented for undoped $M=3$ GNFs, for two values of disorder strength: $w/t=0.10$ and $0.20$, the probability distributions indicate very pronounced negative skewness so that they are strongly asymmetric. The dispersion is also found to rise faster than linearly with the width $w$ of uniform distribution from which on-site energies are taken (unlike the situation for an off-diagonal disorder). Therefore, a long tail of low values of critical field is present. In Fig.~\ref{fig:fig8} we show a dependence of the standard deviation of the critical field for undoped $M=3$ GNF as a function of $w$. Contrary to the case of an off-diagonal disorder, for diagonal disorder $\sigma_{\Delta_{c}}$ increases quadratically with increasing $w$ (see the solid curve fitted to the data). However, the values of critical field dispersion for the same distribution width $w$ are significantly lower for diagonal disorder that for off-diagonal one. In the inset in Fig.~\ref{fig:fig8} a plot of standard deviation of magnetizations of each half of a GNF is included. An analogous quadratic dependence of magnetization on $w$ is visible. 

Let us mention that a similar model of disorder has been used for example in Ref.~\onlinecite{Ezawa2010} to estimate its influence on electronic structure of triangular graphene nanoflakes or in Ref.~\cite{Lee2012} to assess its effect on RKKY interaction. Also in Refs. \cite{Rycerz,Jaworowski2013} the disorder-induced phenomena were discussed for graphene quantum dots.

\section{Final remarks}

In the paper we studied the bow-tie shaped GNFs in external parallel magnetic field with a view on constructing their ground-state phase diagram. Charge-neutral nanoflakes with such a shape are known to exhibit antiferromagnetic ordering of magnetic moments arising at both halves of the structure. We found a field-induced spin-flip transition between antiferromagnetic and ferromagnetic state with a critical field value strongly decreasing with increase of the GNF size. Moreover, we predicted a similar metamagnetic transition between ferrimagnetic and ferromagnetic orientation of magnetic moments at both halves for the case of a GNF doped with a single charge carrier. Critical field values are sensitive to the selection of the Hubbard on-site energy parameter and the transitions themselves do not emerge in absence of coulombic interactions. In addition we studied the influence of diagonal and off-diagonal disorder on the phase diagram. The effect of a diagonal disorder is found to be weaker than of an off-diagonal one. Moreover, the width of distribution of critical fields can be a significant fraction of the average value for the latter case. 

Further developments can include, for example, studying the effect of arbitrary-oriented magnetic field on the properties of the system in question. Another highly interesting issue may be spin-dependent transport properties of bowtie-shaped nanoflakes in magnetic field\cite{Candini}, which gives hope for spintronic applications.

\section{Acknowledgments}

\noindent The computational support on Hugo cluster at Department of Theoretical Physics and Astrophysics, P. J. \v{S}af\'{a}rik University in Ko\v{s}ice is gratefully acknowledged.

\noindent This work has been supported by Polish Ministry of Science and Higher Education on a special purpose grant to fund the research and development activities and tasks associated with them, serving the development of young
scientists and doctoral students.

\end{document}